# Strain induced stabilization of high symmetry phase in MAPbBr$_3$ perovskite

Shankar Dutt[1,2], Joydipto Bhattacharya[3], Kailash Kumar[4], Rajashri Urkude[5], Pankaj R. Sagdeo[4], Archna Sagdeo[1,2,a)]

[1.] *Accelerator Physics and Synchrotrons Utilization Division, Raja Ramanna Centre for Advanced Technology, Indore 452013, India*

[2.] *Homi Bhabha National Institute, Training School Complex, Anushakti Nagar, Mumbai 400094, India*

[3.] *Tyndall National Institute, University College Cork, Cork, Ireland T12R5CP*

[4] *Materials Research Laboratory, Department of Physics, Indian Institute of Technology, Indore, 453552, India*

[5.] *Beamline Development and Application Section, Bhabha Atomic Research Center, Trombay, Mumbai 400085, India*

[a)] Corresponding author: archnaj@rrcat.gov.in

## Abstract

Structural phase transitions in the organic inorganic metal halide perovskites are driven via rearrangement of methylammonium cation and distortion in the PbX$_6$ octahedra. Compositional tuning is usually incorporated for suppression of the structural phase transition in these systems with cation or anion tuning. Along with the compositional tuning, behaviour of strain present in the system can also lead to stabilization of single phase in these systems. In the present investigation, two different samples of CH$_3$NH$_3$PbBr$_3$ perovskite were studied and it is observed that structural phase transition is absent for one of the sample while it is present in the other sample. The non-observance of structural phase transition and stabilization of single phase has been attributed to the perceived tensile strain in the system contrary to the compressive strain observed in the system showing structural phase transitions. This observation is further supported by theoretical calculations. Extended X-ray absorption fine structure measurements revealed distorted octahedra with varying bond lengths along planar and axial directions in both samples, along with observed increase in bond lengths in one of the sample. This stabilization of the cubic phase can enhance device performance and increase overall environmental stability, making these systems more effective for practical applications.

**Introduction**

Organic-inorganic metal halide perovskites[1,2] (OIMHP's) are energy-harvesting materials finding applications in various fields like detectors, solar cell[1,3–5], and other optoelectronic applications[6–9]. These perovskites possess properties like high optical absorption[1], tunable band gap[10], longer carrier diffusion lengths[11], lower charge recombination, defect tolerance etc. OIMHP's based solar cells have exceeded 25% efficiency in less than a decade.[12] Despite all their qualities, these materials unfortunately suffer from poor stability when exposed to heat and moisture[13,14], making their application challenging. However, researchers are working hard to improve the stability of these devices through introduction of passivation mechanisms and multi-layer structures which significantly increase their durability. Moreover, studies on substitution at cation/anion sites are also reported to be affecting chemical stability and other properties of these systems.[15,16,17,18,19] Galvan et al.[20] reported that substitution of mixed halide suppress the structural phase transition in these systems. In a recent study, introduction of $(CH_3)_2N_2^+$ (DMA) cation is shown to stabilize cubic phase in $MAPbI_3$ (MA= $CH_3NH_3$) and photoactive black phase in $CsPbI_3$ which further improves the room temperature performance and stability against water.[21,22,23] High pressure studies are also performed to see structural phase transition behaviour in these systems. Recently, Liang et al.[24] studied high-pressure effects on $MAPbBr_3$ perovskite, identifying phase transitions at 0.8 GPa and 1.8 GPa, with the sequence $Pm\bar{3}m \rightarrow Im\bar{3} \rightarrow Pmn2_1$.

Compositional substitution is well known to perturb structural phases for classical inorganic perovskites which in turn suppress the long range order and dipolar glass type behaviour may appear for the system.[25–27] Simenas et al.[28], recently reported that tuning the A site cation in methylammonium lead bromide, MAPbBr$_3$ with even 14% of dimethylammonium (DMA, (CH$_3$)$_2$NH$_2^+$) suppresses structural phase transition and stabilize cubic phase. Suppression of structural phase transition stabilizes the system as well and hence can improve the performance of the device. Moreover, there are reports of suppression of phase transitions in the materials with presence of strain.[29,30] Presence of strain can influence the stability of different phases, potentially delaying or completely suppressing a phase transition that would occur in the absence of strain.[31,32] Structural phase transitions are well-known and studied characteristic of organic-inorganic metal halide perovskites.[2,33,34] Particularly, MAPbBr$_3$ exhibits four structural phases: a low-temperature orthorhombic phase, a transition to a tetragonal phase (tetragonal 1) around 150K, followed by a transition to a different tetragonal phase (tetragonal 2) near 160K, and finally a transition to a cubic phase around 230K.[33]

The present work primarily focuses on MAPbBr$_3$ perovskite, due to its various advantages over other perovskites in the series, such as enhanced stability[15], an optimal band gap[35], and a wider range of applications[36–38]. In the present manuscript, while exploring the structural phase transitions in MAPbBr$_3$ perovskite system, it was observed that in few samples, phase transition was not observed even upto 5 K. In order to understand the reason behind the non-observance of structural transition in these samples, several samples were prepared and tested. In this investigation, two different samples of MAPbBr$_3$ system were examined, one which shows structural phase transition and one which does not show structural phase transition. Interestingly, it is observed that the structural phase transition could be suppressed even without compositional substitution. The sample that shows structural phase transition is termed as PT and the one that does not show structural phase transition is termed as NPT. The samples were studied using synchrotron X-ray diffraction, Raman spectroscopy, dielectric spectroscopy, extended X-ray absorption fine structure (EXAFS) measurements. Along with experimental investigation, theoretical investigation with density functional theory (DFT) was also performed to support our observations. Detailed analysis of X-ray diffraction data revealed that the presence of tensile strain in the system dominantly suppress structural phase transition. Effect of suppression of phase transition in the sample is also seen in the local coordination behaviour and dipolar behaviour. To the best of our knowledge, for the first time, through EXAFS, direct evidence for the distorted octahedra in the symmetric cubic phase is seen. In contrast to assumed identical bonding strength (Pb-Br bonding) along all the directions in the octahedra, it is observed that the bonding strength differs along the axial and planar directions in the samples.

**Experimental details**

**Synthesis of MAPbBr$_3$ (PT, NPT) sample**

Famous inverse temperature crystallization approach was used to obtain single crystal samples[39]. In order to synthesise $CH_3NH_3Br$ i.e. MABr, 11.3 mL of Hydrobromic acid, HBr and 8.6 mL of methylamine, $CH_3NH_2$ were mixed and stirred in ice bath for around 2-3 hours. Obtained solution was then heated at 60 ºC for 2-3 hours to obtain the precipitate which was kept in desiccator overnight. Later, 2.2394 g of $CH_3NH_3Br$ and 7.3402 g of $PbBr_2$ were mixed and stirred for 3 hours in dimethyl formamide (DMF) to make 20mL, 1M precursor solution of MAPbBr$_3$. After that, few mL's of the solution was heated at 80-85 ºC to obtain cube shaped 1-2 mm size single crystals. Pictures of the as grown samples are shown in Figure S1 of supplementary information file.

**Synchrotron X-ray diffraction**

Temperature dependent X-ray diffraction measurements were carried out at the angle dispersive X-ray diffraction beam line (BL-12), Indus-2 synchrotron radiation source, RRCAT, Indore, India[40] on image plate area detector MAR345dtb with 17 keV photon energy, using Oxford cryostream cryocooler. NIST standard $LaB_6$ was used to calibrate image plate area detector parameters. Heating rate was 6 K/min with a temperature accuracy of ± 0.5 °K. Moreover, synchrotron X-ray diffraction on the PT and NPT samples were taken out with an incident photon wavelength of 0.73012 Å and 0.73007 Å, respectively. XRD data was taken at RT and then cooled to 100 K and then data was taken in heating. Same protocol was followed in the temperature dependant XRD measurements on both the samples. XRD data obtained in the form of rings was converted into intensity vs. two theta pattern using FIT2D software.[40] Lattice parameters were obtained through Rietveld refinement performed using FULLPROF program[41].

**Raman spectroscopy**

Raman spectra were measured using a Horiba LABRAM HR with a 633 nm, Helium-Neon laser, with 0.59 mW power. 633 nm laser was used to avoid any laser induced degradation of the system[42]. Data is collected with a CCD detector working in backscattering mode and a laser band pass filter. Resolution of the setup is 0.3 cm$^{-1}$. Temperature-dependent measurements were done with a Linkam THM600 stage. Cooling rate to reach 93 K from room temperature was 25 degree per minute and the temperature accuracy is ± 0.1 °K.

**Dielectric spectroscopy**

Dielectric measurements were performed using WAYNE KERR 6500B impedance analyser in the temperature range 90–340K and frequency range of 20Hz-5MHz. CRYOCON 22C temperature controller was utilised for temperature measurements. Dielectric data was taken in heating cycle with a heating rate of 1.5 °C with a temperature accuracy of ± 1 °C. The data was taken at every 2 K. Single crystal samples were initially crushed into a fine powder and then pelletized. Silver paste was applied to both surfaces to create contacts, and a parallel plate configuration was employed the dielectric measurements.

**X-ray absorption fine structure**

XAFS (XANES and EXAFS) measurements at Pb $L_3$-edge (13035 eV) were performed at scanning EXAFS beamline (BL-09) at Indus-2, synchrotron radiation source, RRCAT, India. The measurements were carried out in transmission mode. The energy calibration was performed using Pb foil as a reference. Single crystals were crushed into fine powder and which was then mixed uniformly with cellulose matrix and pressed into thin pellets of 10 mm diameter. These pellets were then sandwiched between two Kapton adhesives tapes, for the measurements. The XAFS spectra of the Pb $L_3$ edge was collected in the energy range between 12925 eV to 13400 eV only due to the presence of Br K edge (13474 eV) after that. ATHENA and ARTEMIS software were used to analyse the data for normalizing and fitting the shells.[43,44] The standard normalization of the data and background subtraction procedures were followed using the ATHENA software version 0.9.26 to obtain normalized XANES spectra. This involved preliminary reduction of the EXAFS raw data, background removal of the X-ray absorption data $\mu(E)$, conversion of $\mu(E)$ to $\chi(k)$, normalization and weighting scheme, using AUTOBK[45]. Detailed discussion of XAFS analysis are provided in supplementary information file.

**Results and discussions**

**Synchrotron X-ray diffraction studies**

As discussed in the introduction, two samples were selected for the present investigation, one that shows structural phase transition (PT) and other that does not (NPT), as a function of temperature. Room temperature X-ray diffraction pattern for both the samples are shown in figure 1(a). It can be observed that all the peaks for both the samples correspond to the perovskite structure of $MAPbBr_3$ and presence of no extra peaks confirms the phase purity of the prepared samples. It is clear from the figure that the XRD patterns of the two samples (PT and NPT) appears similar. Further, for investigating the structural phase transition in the two samples, XRD patterns were collected at 100 K for both the samples, where MAPbBr$_3$ system is expected to be in orthorhombic phase. The obtained temperature dependent synchrotron X-ray diffraction (SXRD) pattern for the two samples at 100K is shown in figure 1(b). At 100 K, peak splitting and emergence of many other peaks can be clearly seen in figure 1(b) for the PT sample. These extra peaks and peaks splitting is attributed to the uplifting of degeneracy in orthorhombic phase. Inset of figure 1(a) and 1(b) shows schematic of cubic and orthorhombic structure of MAPbBr$_3$, respectively at RT and 100 K as obtained through VESTA software[46]. A complete temperature dependent X-ray diffraction with structural phase transition from orthorhombic to tetragonal and then to cubic a part of other publication[47].

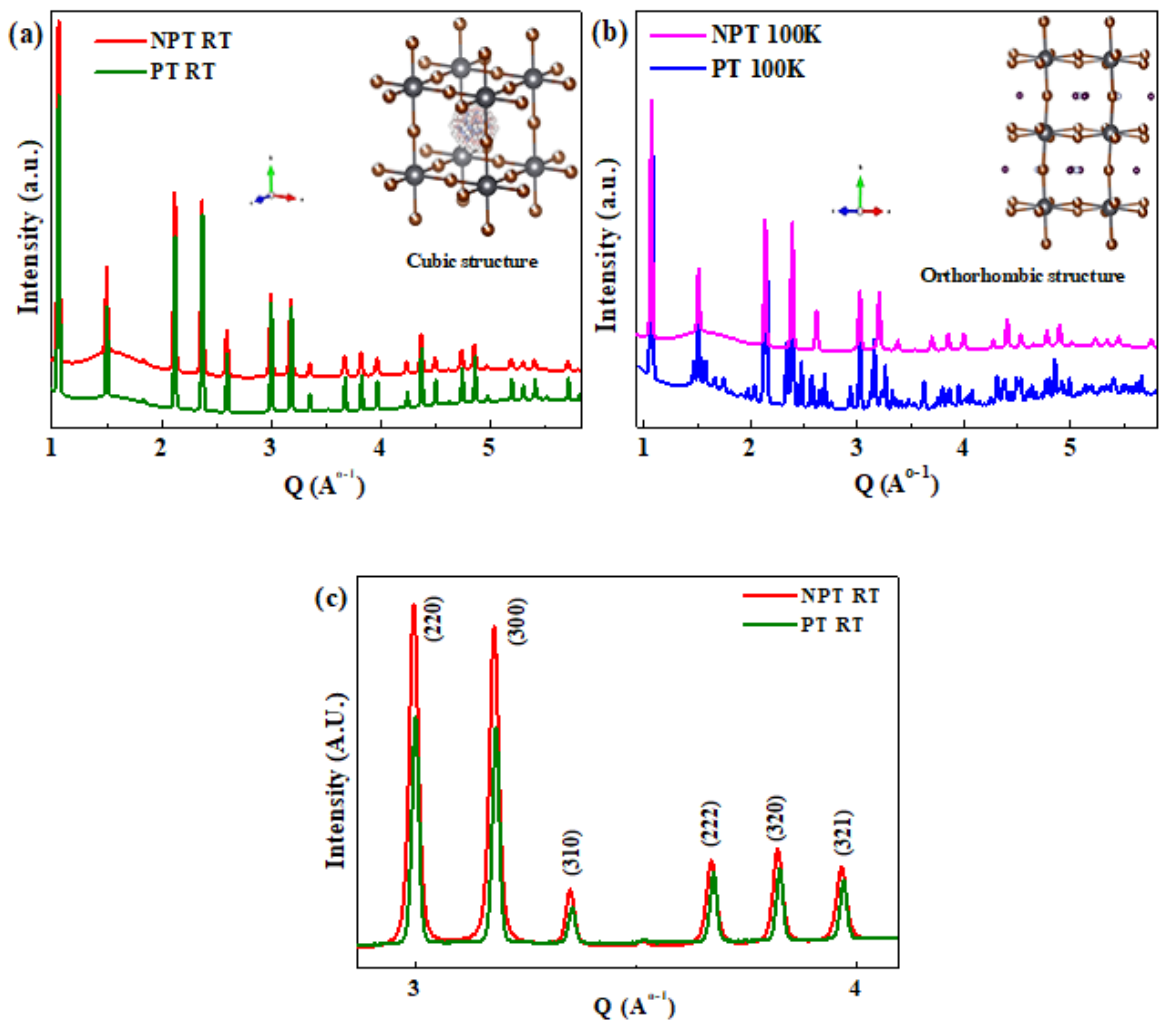

*Figure 1. X-ray diffraction data at (a) Room temperature and (b) 100 K for both the samples. Inset of (a) shows structure of MAPbBr$_3$ in cubic structure at RT (obtained from refinement of diffraction data of NPT at RT) while inset of (b) shows orthorhombic structure (obtained from refinement of diffraction data of PT at 100 K) and (c) shows a visualization of selected Bragg reflections from the two samples at RT.*

However, X-ray diffraction data obtained at 100 K for NPT sample clearly shows that it remains in the cubic structure even at 100 K, where it is expected to be in orthorhombic structure otherwise. X-ray diffraction measurements have been performed at 5 K also, in order to see the structural phase transition, in case it has been shifted to lower temperatures, however, no phase transition was observed even up to 5 K.

Reitveld refinement for estimation of lattice parameters has also been performed on the powder XRD data of two samples, using Pnma space group for orthorhombic phase and Pm$\bar{3}$m for cubic phase, with the help of Full Prof program. Further details of fitting are provided in supplementary file. Obtained values are found to be in agreement with the literature.[48,35] Bond length between Pb and Br atom, for both cubic as well as orthorhombic structures were also extracted from the structure obtained using VESTA software. Results obtained from the diffraction data are tabulated in table 1.

*Table 1. Parameters obtained through temperature dependent X-ray diffractions studies.*

| Sample | Lattice parameter (100 K), ( Å ) | Lattice parameter (RT), ( Å ) | Pb-Br bond length ( Å ) |
|---|---|---|---|
| MAPbBr$_3$ (PT) | a=7.979 (±0.006), b=11.838 (±0.007), c=8.569 (±0.007) | 5.924 (±0.002) | 2.9624 (±0.0005) |
| MAPbBr$_3$ (NPT) | a=5.883 (±0.001) | 5.930 (±0.001) | 2.9651 (±0.0001) |

Thus, through the structural characterization, it is observed that the structural phase transition seems to have suppressed in the NPT sample. It exhibits cubic phase at RT as well as at 100

K. Whereas, PT sample exhibits orthorhombic phase at 100 K and a cubic phase at RT. Further, to examine the changes in local bonding behavior between the two samples, vibrational spectroscopy was carried out.

**Raman spectroscopy**

Raman is a well-known technique to study local vibrational behaviour and structural phase transitions of any system[2,49]. Raman spectra taken for both the samples at room temperature and 93 K is shown in figure 2. Low frequency modes (0-200cm$^{-1}$) are attributed to the motion of heavier inorganic part i.e. $PbBr_6$ octahedra, while the higher frequency modes are related to lighter organic part ($CH_3NH_3$). Low frequency modes show dominant changes in case of structural phase transition. It is observed that a large number of Raman modes appear at 93K for the PT sample as compared to room temperature Raman spectra. Appearance of many modes at low temperature is due to symmetry lowering in the orthorhombic phase[2,50]. However, for the NPT sample, the Raman spectra is almost similar for room temperature and lowest temperature (93 K). Higher frequency modes related to vibration of organic part and its interaction with the inorganic octahedral cage does not show very significant changes around the structural phase transition. Observed Raman spectra align well with the literature[42,51].

A recent report by Talit et al.[52] discuss strain effect on the vibrational spectra of one of the perovskites of this family i.e. $MAPbI_3$. With the help of density functional theory, they have predicted Raman mode frequencies with the external compressive and tensile strain compared to zero strain condition. They have observed that low frequency modes are observed at higher frequency in case of compressive strain compared to the case of tensile strain while the mid frequency Raman modes are not good to probe strain effects due to $CH3NH3^+$ rotation. To compare with the results of Talit et al.[52] and to obtain the Raman shift behaviour for the two samples at RT and 93 K, Raman spectra shown in figure 2 was fitted with a lorentzian peak profile. Results of the present work also align with their observation i.e. for low frequency modes, compressive strain blue shifts the modes as compared to tensile strain while for higher frequency, they are red shifted for compressive strain with respect to tensile strain. Obtained band frequencies from the Raman data in the present work are tabulated in table P5 supplementary file.

In addition, recent report by Zhang e*t al.*[53] need to be mentioned, where they have studied the effect of strain on electronic structure of MAPbI$_3$. Zhang and group have also used density functional theory to study the behaviour. They have observed that on increasing the compressive strain in the system, system undergoes a structure change from tetragonal to orthorhombic phase. However, on increasing tensile strain, system remains in the tetragonal phase, an observation aligning with the results in the present study.

Motion of methylammonium cation in hybrid perovskites is known to be responsible for structural phase transitions. MA cation has a permanent dipole due to the permanent charge associated with it, this makes the dipolar study of these behaviour rather interesting and to investigate that, dielectric measurement on these two samples have been performed.

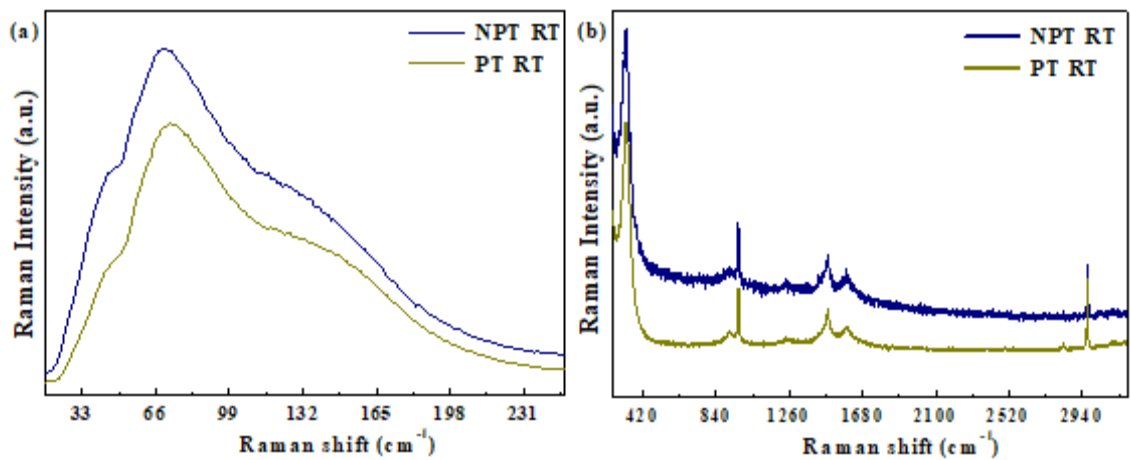

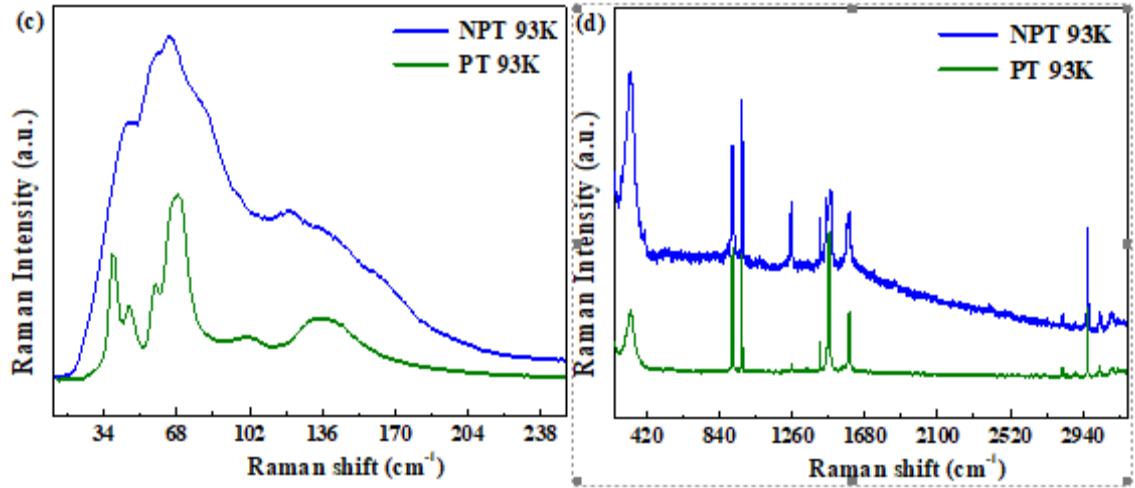

*Figure 2. Raman spectra of PT and NPT sample at RT of (a) low frequency modes (0-250 cm$^{-1}$), (b) high frequency modes (250 cm$^{-1}$ and higher) and at 93 K of (c) low frequency and (d) high frequency Raman modes.*

**Dielectric spectroscopy**

Dielectric measurement can be helpful in understanding dipolar nature of a system. Thus, dielectric spectroscopy as a function of temperature is performed on the two samples to explore the phase transitions ( if at all it exists) and dipolar dynamics in the frequency range of 20 Hz-5 MHz. Results obtained from temperature dependent dielectric spectroscopy for selected frequencies are plotted in figure 3(a) and 3(b). Inset of plots shows dielectric loss plotted as a function of temperature for both the samples. For PT sample, a sharp increase in the real part of the dielectric permittivity is seen at 145 K corresponding to the structural phase transition from low temperature orthorhombic structure to tetragonal phase. The sharp rise is attributed to dominant contribution of the MA dipoles to dielectric constant. This results matches well with reported literature[54,55]. Sudden kinks seen around 200 K and 250 K in figure 3(a) are due to pouring of liquid nitrogen into the measurement chamber.

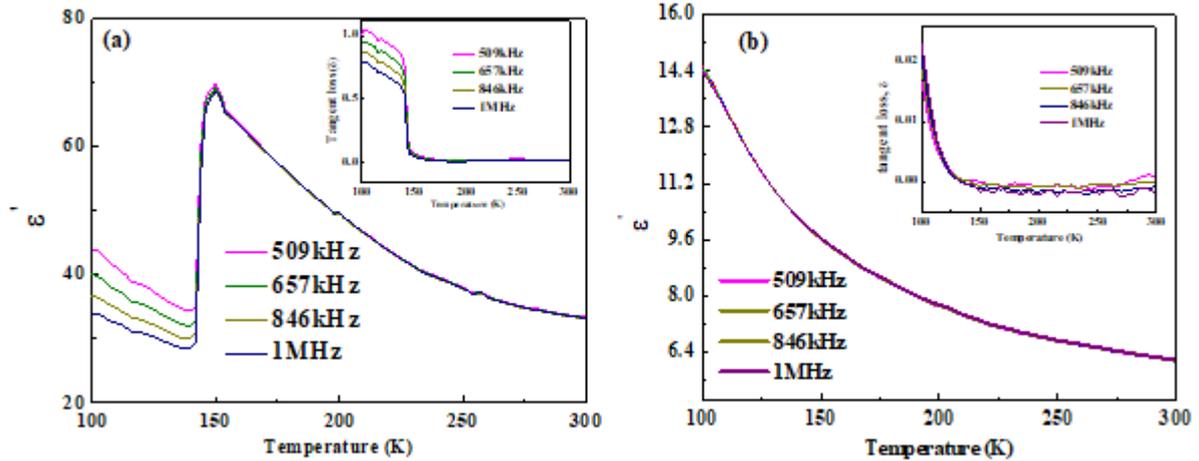

*Figure 3. Dielectric permittivity and loss tangent plotted against temperature for the (a) PT sample and (b) NPT sample.*

Dielectric data of the NPT sample is shown in figure 3(b) where in the inset of the plot, temperature dependent dielectric loss is shown. No anomaly is seen in the dielectric data of NPT sample. Absence of any anomaly in 100-300 K indicates that dipolar ordering is not changing in the studied temperature range for NPT sample which further suggests absence of structural phase transition in NPT. It is seen that while in the PT samples, there are clear anomalies around the structural phase transition temperature, indicative of existence of different structural phases over the temperature range. However, in the NPT sample, no such anomaly is observed, further suggesting the freezing of a single phase for the entire studied temperature range. The observation supports the findings of diffraction measurement where the high temperature cubic phase was seen to freeze even for the lowest temperature (100K) for the NPT sample.

From the literature, it is well known that the presence of micro strain in samples affects the structural phase transition behaviour in systems.[32,31] Thus, it is suspected that the behaviour of micro strain present in the system might be the possible reason for the observed behaviour in the two systems. To investigate the micro strain present in the two samples, measured XRD data was analysed thoroughly. Apart from this in continuation with the Raman results, to explore the type of strain, whether compressive or tensile, present in the system, X-diffraction is the best tool. This information can be extracted using the peak profile analysis of the x-ray diffraction data. Thus, to investigate the micro strain and hence the type of strain present in the two samples, measured XRD data was analysed thoroughly.

**Strain analysis**

It is well known that strain could modulate the orientation and motion of organic cation and polyhedral, and it is suspected that the freezing of structural phase transition in the system might be arising from the microstrain present in the material[56,57]. To see the behaviour of microstrain present in the two samples, along with their crystallite size, Williamson-Hall equation[58] is used which is given by equation 1.

$$\beta cos\theta = 4\varepsilon sin\theta + \frac{0.9\lambda}{D} \quad (1)$$

Here $\beta$ is the full width at half maximum (FWHM) value of the diffraction peak, $\varepsilon$ corresponds to average strain-induced in the sample, $\theta$ is the diffraction angle, $\lambda$ is the wavelength of X-rays and D is average crystallite size. Thus, Williamson-Hall (WH) plot was obtained for RT and lowest temperature SXRD data, for both the samples. For that, FWHM values of the diffraction peaks were calculated and $\beta cos\theta$ Vs $4sin\theta$ has been plotted, as can be seen in figure 4. As W-H equation is similar to equation of a straight line, the obtained plot is fitted using the straight-line equation. The slope of the fitted straight line provides the micostrain present in the samples and the average crystallite size is evaluated through y intercept. It can be clearly seen from figure 4(a), which shows the W-H plots for PT sample, that at room temperature, the slope of the W-H plot for PT sample is negative. The negative slope indicates that the strain present in the system is compressive in nature, while positive slope indicates tensile strain in the system[59,60]. For comparison, W-H plot has also been obtained for the XRD data at 100 K for NPT sample and has been shown in figure 4(c). The values of micro strain and crystallite size calculated using W-H plot, for both the samples at RT and NPT sample at 100 K are given in Table-2.

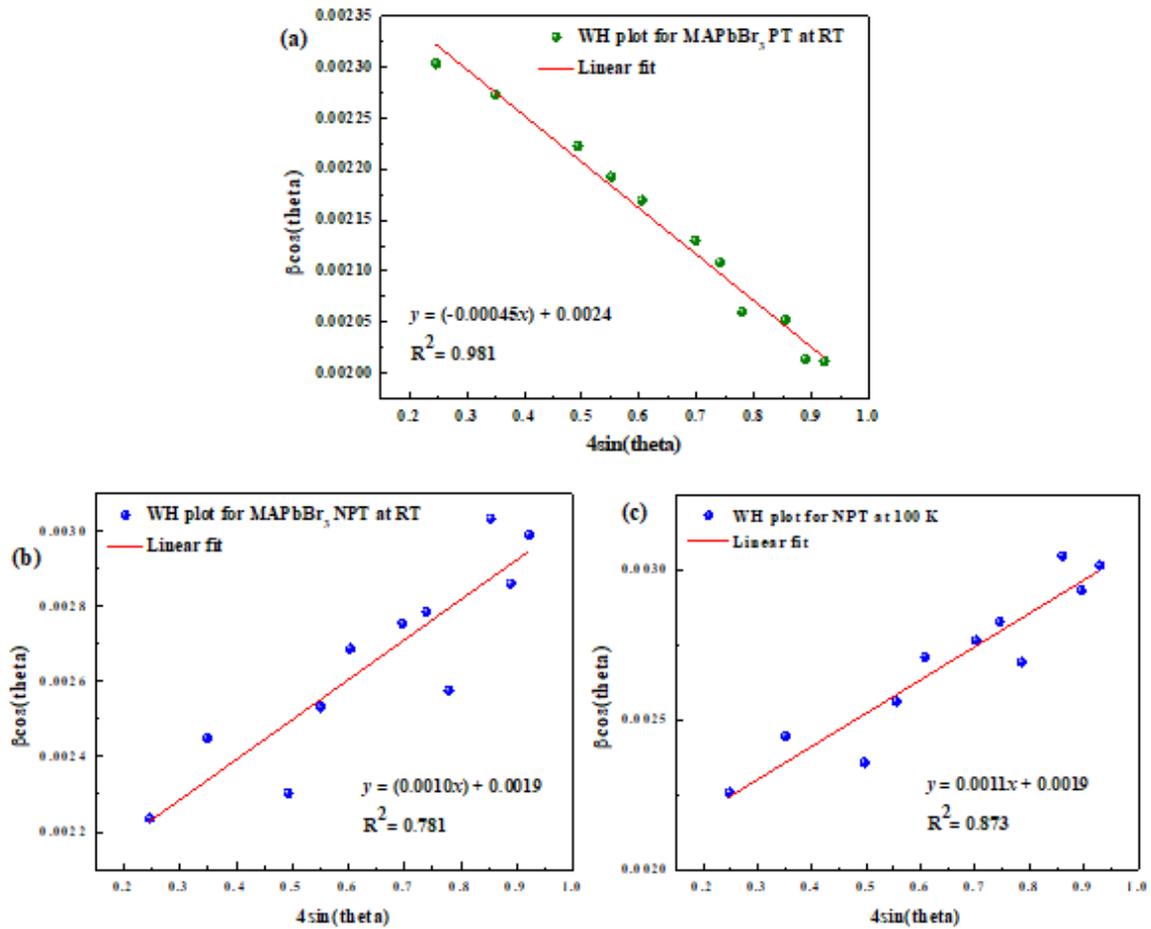

*Figure 4. Williamson-Hall plot for (a) room temperature of PT and (b) room temperature of NPT, (c) 100 K of NPT data*

*Table 2. Parameters obtained through temperature dependent X-ray diffractions studies.*

| Sample | Crytallite size (nm) (100 K) | Crytallite size (nm) (RT) | Strain (100 K) | Strain (RT) |
|---|---|---|---|---|
| MAPbBr$_3$ (PT) |  | 328.5 ($\pm 0.14$) |  | $-0.4 \times 10^{-3}$ ($\pm 0.01 \times 10^{-3}$) |
| MAPbBr$_3$ (NPT) | 333.5 ($\pm 1.52$) | 333.5 ($\pm 2.01$) | $1.1 \times 10^{-3}$ ($\pm 0.17 \times 10^{-3}$) | $1 \times 10^{-3}$ ($\pm 0.17 \times 10^{-3}$) |

Thus, from W-H plot for the NPT sample, a tensile strain of $1\times10^{-3}$ was obtained at RT which increases to $1.1\times10^{-3}$ at 100 K, whereas, the crystallite size is found to be similar for RT (333.5 nm) and 100K (333.5 nm) temperature. It is seen that strain and crystallite size does not vary much for the NPT sample even with the huge temperature difference between RT and 100 K. This is another consequence of freezing of cubic structure throughout the temperature region for NPT sample. Thus, from the entire study, the only difference in the two samples at RT is the different behaviour of strain present in the two samples (NPT and PT) which is speculated to be the reason for the drastically different structural transition behaviour in the two samples. Presence of tensile strain might be breaking the ordering in the system which is responsible for reorientation of MA cation and thus driving the phase transition on temperature variation. The microstrain could have been induced in the system during the synthesis procedure which involves complex chemical reactions.

Report by Zhou *et al.*[61] excellently discuss strain engineering in Formadinium based perovskites where internal and external strain have been discussed and various techniques to study strain and tune it are also discussed. Moreover, Murali[62] and group have studied the effect of annealing temperature on strain generated in the system for $MAPbBr_3$. They have observed that with the variation of synthesis temperature, the amount and nature of the strain also varies. It is seen that for the temperature range of 60±5 ºC, perovskite system exhibits relaxed cubic structure and on both sides of this temperature range, structure gets distorted by strain. Lower annealing temperature than this have tensile strain while higher temperature have compressive strain. The amount of strain in the present investigation for the temperature range of 60-90 ºC is comparable to the values reported by them.

These observations were further validated with the help of theoretical calculations on this the system in accordance with the previous studies[63,64]. Main findings of density functional theory (DFT) results are (a) For compressive strain, orthorhombhic phase was observed to be more stable, while for tensile strain cubic phase was found to be more stable and (b) Tilting of $CH_3NH_3$ cation in the cubic phase under tensile strain leads to a lowering of energy. A comprehensive discussion of the theoretical study, performed using density functional theory (DFT), is provided in the supplementary information file.

Further, to investigate the local structural behaviour in the two systems, XAFS measurement were performed.

**XAFS studies (X-ray absorption fine structure)**

EXAFS analysis is valuable for exploring local structural changes and structural disorder in a system. In perovskite materials, these structural fluctuations play a crucial role in influencing their outstanding optoelectronic properties. Advantage of using EXAFS is that it provides a local picture of the probed atom as compared to X-ray diffraction where an average picture of the structure is obtained. In an important study on $MAPbBr_3$ perovskite, Weadock and group[65] studied temperature dependence of local disorder in the $MAPbBr_3$ system using EXAFS and X-ray diffraction along with *ab initio* molecular dynamics (AIMD) across the structural phase transitions. It is observed that across the phase transition of $MAPbBr_3$, Pb-Br effective spring constant shows an anomalous increase attributed to the redistribution of the charge density via affecting the Br…H bonding. Moreover, it is observed that the structural disorder in the $MAPbBr_3$ system is majorly due to thermally activated anharmonic dynamics rather than static disorder. Another report by Bridges and group[66] explored structure, bonding and asymmetry in $APbBr_3$ (A=$(NH_2)_2CH$=FA, Cs and $CH_3NH_3$) systems via temperature dependent EXAFS study on Pb $L_3$-edge and Br K-edge. Static disorder was found to be increasing with the increase in the A site cation (for $MA^+$ and $FA^+$) compared to smaller value in $CsPbBr_3$. Spring constant values in the three systems are found to be comparable lying in the range from 1.2 to 1.95 eV/ $Å^2$.

Thus, local information becomes crucial for better understanding of the system's physical properties. For that, XAFS comprising of X-ray absorption near edge spectroscopy (XANES) and Extended X-ray absorption fine structure (EXAFS) measurements on the two samples at Pb $L_3$-edge have been done at room temperature.

It is observed that although the absorption edge i.e. the XANES part, of the two samples does not show any noticeable changes (as can be seen in Figure S6 of supplementary) but the EXAFS region indicated interesting behaviour. The first shell, for the Pb $L_3$-edge, gives the information of Pb-Br bond length of $PbBr_6$ octahedra. The local structure around the Pb atom in NPT and PT have been investigated for the any local changes in structure**or** presence of any disorder, which is giving rise to the difference in two samples, by probing Pb $L_3$-edge. However, the difference between the energy positions of Pb $L_3$ edge (13035 eV) and Br K edge (13474 eV) is very close (i.e., energy difference is of only 439 eV) due to which extended regions of the Br K-edge interfere with Pb $L_3$ edge which limits the usable k range. Due to this, fits were performed in the $R$ -space for nearly (1 - 5) Å range and in k and $q$ - space for nearly (3 – 9) Å$^{-1}$ range.

The normalized EXAFS spectra associated with Pb $L_3$ edge for NPT and PT samples are as shown in figure 5 (a-d) in R space. Data processing and analysis has been done as per the discussion given in the experimental part and is also given in supplementary information file. The first peak with maximum magnitude in figure 5 corresponds to the first coordination shell having six neighbouring Br atoms around the absorbing Pb atoms in NPT and PT sample.

EXAFS data fitting was started with the with the theoretical model based on the crystal structure parameters obtained using analysis of X-ray diffraction data of PT and NPT samples, with space group (Pm3̄m)[65,66]. Figure 5 (a-h) shows the Pb $L_3$ edge EXAFS spectra fitted in R and k spaces, respectively.

Two different structural models were employed to fit the EXAFS data of PT and NPT sample. In first model, the data was fitted using scattering paths obtained from the Pm3̄m structure having 6 equal scattering lengths between Pb and the 6 Br atoms that forms the octahedra. While in the second model, 2 + 4 (two axial path lengths and 4 planer path lengths in the octahedra) model was used. The motivation for using the latter model was to improve the quality of fit which is parameterized through R factor and the possibile existence of octahedral distortion[67]. Results obtained from the fittings of the EXAFS equation with experimental data in *R* and k spaces, respectively are shown in figure 5(a-h) for PT and NPT sample. Table 3 summarizes the fitting parameters obtained from EXAFS data analysis. It can be seen from the figures that experimental data matches well with the theoretical curves calculated for this model. From the first coordination shell fitting obtained through first model, it is observed that Pb atom is coordinated to six Br atoms at ~2.939 Å in PT and ~2.947 Å in NPT, bond distance for the first coordination shell. While for the second model, fitting yields two axial Pb-Br bond distances at ~2.617 Å in PT and ~2.899 Å in NPT and other four planar Pb-Br bonds at ~2.865 Å in PT and ~2.736 Å in NPT, respectively. It is worth noting that the static disorder as revealed by $\sigma^2$ for the first coordination shells of PT and NPT samples are almost similar, but the reliability of the fit i.e. the R factor is better in case of second model.

Thus, it is observed from the table 3 as well as figure 5 (a-h) that for both the samples, better fitting was observed while considering the second model where two types of scattering path between Pb atom and Br atoms is considered, indicating that in both the samples, the octahedra is not regular but distorted. Also, the scattering path lengths i.e. the bond length obtained by using the first model in NPT sample is larger as compared to PT sample.

Whereas, if second model is considered than the axial bond length is greater for NPT sample as compared to PT sample. While the planar bond lengths are smaller for NPT sample as compared to the PT sample. However, the volume of the octahedral for the NPT sample is greater than the volume of the PT sample. Thus, it is observed that while the distortion in the octahedra is present in both the samples, giving rise to strain in each, but the nature of the strain differs. In the PT sample, the smaller octahedral volume results in compressive strain, which compresses the system. Conversely, NPT sample, the larger volume of the octahedra leads to tensile strain, which expand the system.

Thus, it is observed that the octahedra in both PT and NPT samples are distorted. The octahedral volume is greater in case of NPT as compared to PT sample, indicating a weaker bonding for the case of NPT relative to PT which in turn is expected to affect the interaction dynamics between $PbBr_6$ and organic cation to freeze the structural phase transition.

The increased bond distances observed in the NPT sample are anticipated to influence the reorientation and tilting of the octahedra, as well as their interactions with the methylammonium cation. These interactions are responsible for driving the structural phase transition in these systems. As a result, the increased bond distances are expected to modify these dynamics in such a way that the structural transition is suppressed, effectively freezing the phase transition within the system.

Thus, it is contemplated that the presence of tensile strain the NPT sample is arresting the expected phase transition in the sample. It is speculated that different nature of the strain is suspected to be arising from the processes involved in the synthesis, which however, as of now, remains an open question to the research community and can help in paving the way for future work in this field. From the present study, it is speculated that by carefully understanding and optimizing synthesis procedure, a single phase can be stabilized in these systems. Moreover, the possibility of stabilizing symmetric cubic phase throughout helps in better stability and performance of the material for various applications[21,22].

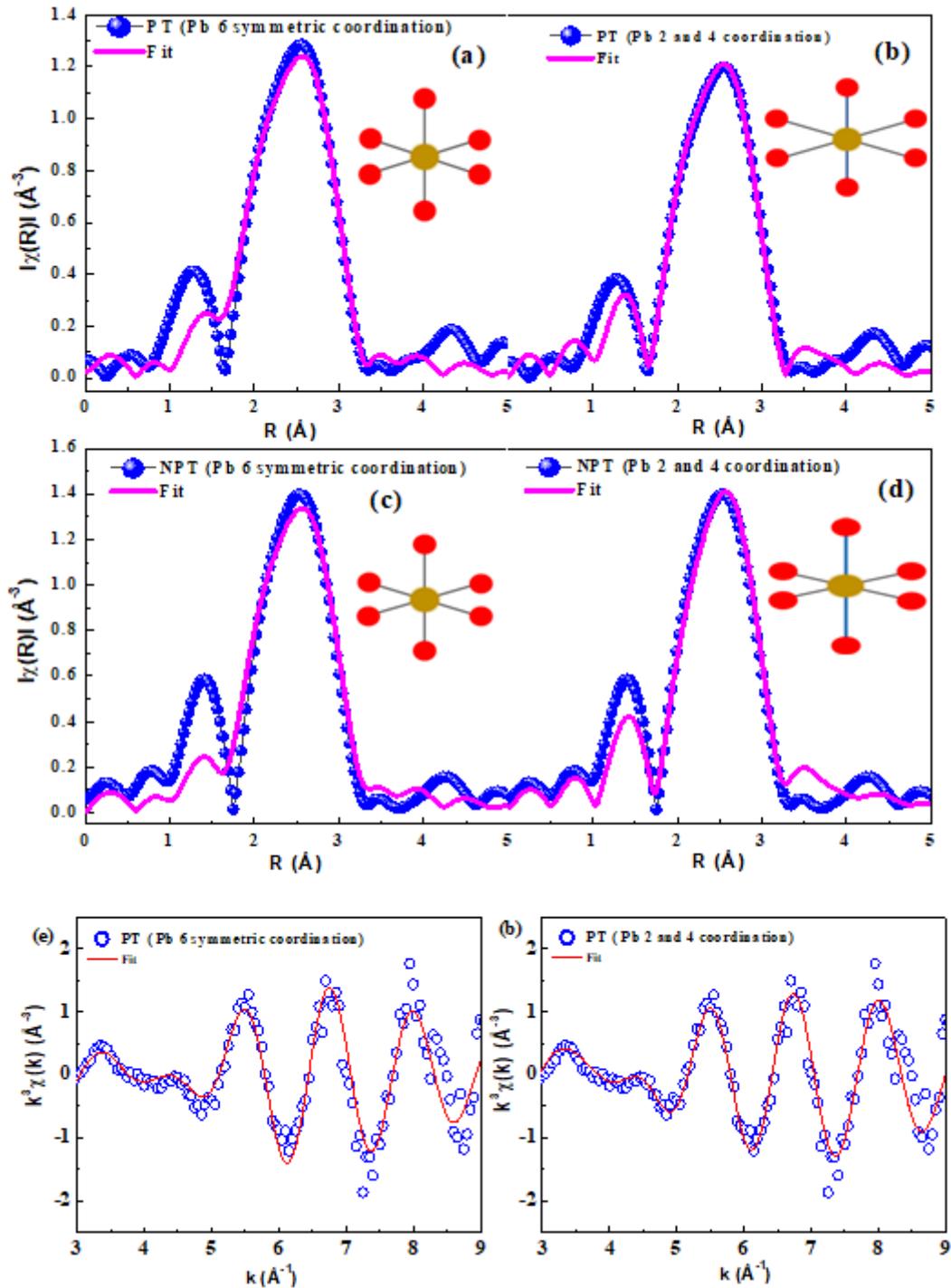

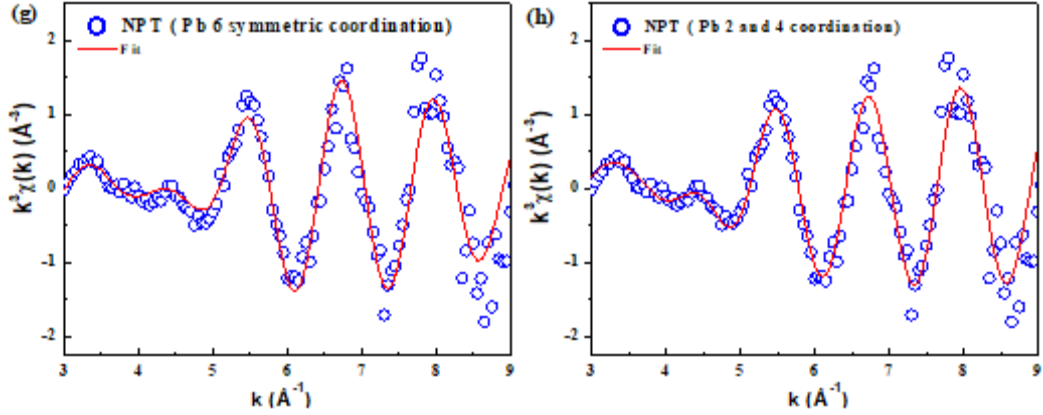

***Figure 5.*** *Pb L3 edge FT-EXAFS spectra for (a,e) PT 6 equal coordination, (b,f) PT two type of coordination (2 and 4), (c,g) NPT 6 equal coordination and (d,h) NPT two type of coordination (2 and 4) in R and k space, respectively.*

***Table 3.*** *Results obtained from the fitting of EXAFS data of Pb $L_3$-edge of the samples.*

| Sample | Bond | N | R (Å) | $\sigma^2$ | $\Delta E_0$ | R-factor |
|---|---|---|---|---|---|---|
| PT | Pb-Br | 6 | 2.939 +/- 0.024 | 0.022 +/- 0.004 | 0.234 +/- 3.005 | 0.060 |
|  | Pb-Br1 | 2 | 2.617 +/- 0.028 | 0.029 +/- 0.003 | -8.602 +/- 3.067 | 0.021 |
|  | Pb-Br2 | 4 | 2.865 +/- 0.025 | 0.027 +/- 0.004 |  |  |
| NPT | Pb-Br | 6 | 2.947 +/- 0.024 | 0.019 +/- 0.003 | 0.371 +/- 3.159 | 0.064 |
|  | Pb-Br1 | 2 | 2.899 +/- 0.027 | 0.015 +/- 0.004 | -9.754 +/- 4.317 | 0.029 |
|  | Pb-Br2 | 4 | 2.736 +/- 0.052 | 0.032 +/- 0.007 |  |  |

*N*: coordination number; *R*: bond distance; $\sigma^2$: Debye-Waller factor; $\Delta E_0$: the inner potential correction. *R* factor: goodness of fit.

**Conclusions**

In the present work, structural phase transition behaviour in two different set of MAPbBr$_3$ perovskite samples, have been investigated. Synchrotron powder X-ray diffraction measurement indicated a clear absence of the structural phase transition in NPT sample as a function of temperature, as compared to PT sample. Diffraction data showed stabilization of cubic phase even at lowest temperature (100 K) for the NPT sample. The result was supported by Raman spectra as well as dielectric permittivity measurements. A thorough XRD analysis via Williamson-Hall plots shows the presence of tensile strain in the sample showing no structural transition. Consistent with the micro strain studies through SXRD,

EXAFS study showed a possibility of different scattering lengths along planar and axial bonds in PbBr$_6$ octahedra. The bond lengths in the NPT sample is observed to be relatively higher compared to PT sample which might be giving rise to the tensile nature of strain in the NPT sample. Weaker interaction between the organic and inorganic part in the NPT sample might be responsible for the presence of tensile strain in the system, leading to the stabilization of the cubic phase throughout. Theoretical study also suggests that orthorhombic structure is more energetically favourable in the case of compressive strain as compared to cubic phase while cubic phase is energetically more favourable in the case of tensile strain, supporting the experimental findings. It is suspected that the difference in both the samples arise during the synthesis procedure which require further exploration. Understanding and enhancing the stability and functionality of the related devices depends heavily on these aspects of hybrid perovskites. It would be very interesting to see how the stabilization of single phase without compositional tuning affects the performance of device. Moreover, stabilization of the single phase is expected to improve overall stability of these systems.

**Acknowledgements**

The authors would like to thank U. D. Malshe, Director RRCAT, Dr. Tapas Ganguli, Head APSUD for their encouragement and constant support. The authors acknowledge the Department of Science and Technology Fund for Improvement of S&T Infrastructure (SR/FST/PSI-225/2016) for providing funding for the Raman spectrometer. J. B. thanks Irish Research Council Laureate Advanced Award Programme (Project ID: IRCLA/2023/1842) for the funding. S. D. is thankful to Homi Bhabha National Institute and RRCAT for providing research fellowship.

**Author contributions**

**Shankar Dutt:** Formal analysis, Data curation, Sample synthesis, sample characterization, data analysis, problem identification, manuscript, Conceptualization, writing, and modification. **Joydipto Bhattacharya:** DFT investigations and rigorous discussion. **Rajashri Urkude:** XANES and EXAFS measurement and analysis. **Kailash Kumar:** Raman measurements. **Pankaj R. Sagdeo:** Raman measurement and fruitful discussions. **Archna Sagdeo:** Manuscript, Conceptualization, Supervision of scientific problems, rigorous discussions and fruitful suggestions.

**Conflict of Interest**

The authors declare that they have no conflict of interest.

**Data availability**

Data will be made available on request.